\documentclass[10pt,conference]{IEEEtran}

\usepackage{balance} 
\usepackage{subcaption}

\usepackage{graphicx} 
\usepackage{xcolor}
\usepackage{float}
\usepackage{cite}
\usepackage{amsmath,amssymb}
\usepackage[outdir=./]{epstopdf}
\usepackage{afterpage}
\usepackage{optidef}
\usepackage{orcidlink}
\usepackage{soul}
\usepackage{tikz}
\usetikzlibrary{spy}
\usepackage{pgfplots}
\usepackage{pgfplotstable}
\usetikzlibrary{shapes.geometric, fit, arrows, positioning, backgrounds}
\usetikzlibrary{calc}
\usepackage{multirow}

\usepackage{comment}
\usepackage{optidef}

\usepackage{colortbl}  
\usepackage{tcolorbox} 
\usepackage{algorithm}
\usepackage{algpseudocode}
\usepackage{bm}
\usepackage{dsfont}

\DeclareMathOperator*\argmax{arg\,max}

\usepackage{xspace}

\usepackage{hyperref}
 \hypersetup{
     pdfpagemode=UseNone,
     pdfpagelabels=false,
     bookmarks=false,
    colorlinks=false,
 	linkbordercolor=white,
 	urlbordercolor=white,
 	pdfborder={0 0 0}
 }  
 
 \usepackage{acro}

\usepackage{ifthen}

\acsetup{
    first-style = long-short
}

\let\oldacs\acs
\let\oldacf\acf

\makeatletter
\RenewDocumentCommand{\ac}{m}{%
  \@ifundefined{myac@#1@seen}{%
    \global\@namedef{myac@#1@seen}{}%
    \leavevmode
    \raisebox{1em}{\hypertarget{#1-first}{}}%
    \oldacf{#1}%
  }{%
    \hyperlink{#1-first}{\oldacs{#1}}%
  }%
}
\makeatother

\DeclareAcronym{ATC-FWI}{
    short = ATC-FWI,
    long  = Adapt-Then-Combine Full Waveform Inversion
}
\DeclareAcronym{FWI}{
    short = FWI,
    long  = Full Waveform Inversion
}
\DeclareAcronym{NN}{
    short = NN,
    long  = Neural Network
}
\DeclareAcronym{ML}{
    short = ML,
    long  = Machine Learning
}
\DeclareAcronym{JSCC}{
    short = JSCC,
    long  = Joint Source and Channel Coding
}
\DeclareAcronym{FL}{
    short = FL,
    long  = Federated Learning
}
\DeclareAcronym{OAC}{
    short = AirComp,
    long  = Over-the-Air Computation
}
\DeclareAcronym{ELBO}{
    short = ELBO,
    long  = Evidence Lower BOund
}
\DeclareAcronym{MILBO}{
    short = MILBO,
    long  = Mutual Information Lower BOund
}
\DeclareAcronym{dELBO}{
    short = dELBO,
    long  = Distributed Evidence Lower BOund
}
\DeclareAcronym{dMILBO}{
    short = dMILBO,
    long  = Distributed Mutual Information Lower BOund
}
\DeclareAcronym{AWGN}{
    short = AWGN,
    long  = Additive White Gaussian Noise
}
\DeclareAcronym{DNN}{
    short = DNN,
    long  = Deep Neural Network
}
\DeclareAcronym{SGD}{
    short = SGD,
    long  = Stochastic Gradient Descent
}
\DeclareAcronym{MSE}{
    short = MSE,
    long  = Mean Square Error
}
\DeclareAcronym{MAS}{
    short = MAS,
    long  = multi-agent system
}
\DeclareAcronym{CNN}{
    short = CNN,
    long  = Convolutional Neural Network
}
\DeclareAcronym{NMSE}{
    short = NMSE,
    long  = Normalized Mean Square Error
}
\DeclareAcronym{SSIM}{
    short = SSIM,
    long  = Structural Similarity Index Measure
}
\DeclareAcronym{ViT}{
    short = ViT,
    long  = Vision Transformer
}
\DeclareAcronym{ResNet}{
    short = ResNet,
    long  = Residual Network
}
\DeclareAcronym{PReLU}{
    short = PReLU,
    long  = Parametric Rectified Linear Unit
}
\DeclareAcronym{PSNR}{
    short = PSNR,
    long  = Peak Signal to Noise Ratio
}
\DeclareAcronym{SNR}{
    short = SNR,
    long  = Signal to Noise Ratio
}
\DeclareAcronym{pdf}{
    short = PDF,
    long  = Probability Density Function
}
\DeclareAcronym{UMAP}{
    short = UMAP,
    long  = Uniform Manifold Approximation and Projection
}
\DeclareAcronym{t-SNE}{
    short = t-SNE,
    long  = t-distributed Stochastic Neighbor Embedding
}
\DeclareAcronym{HDBSCAN}{
    short = HDBSCAN,
    long  = Hierarchical Density-Based Spatial Clustering of Applications with Noise
}

\DeclareMathAlphabet\mathboldsf{OT1}{cmss}{bx}{n}

\newlength\figH
\newlength\figW
\newlength\figWsmall
\newcommand*{\nth}[2]{
    \foreach \xindx [count=\kcount from 0] in #1 {
        \ifnum\kcount=#2
            \xindx
        \fi
    }
}

\newboolean{xtickAutomatic}
\setboolean{xtickAutomatic}{true}

\newboolean{ytickAutomatic}
\setboolean{ytickAutomatic}{true}

\newboolean{xMinAutomatic}
\setboolean{xMinAutomatic}{true}
\newboolean{xMaxAutomatic}
\setboolean{xMaxAutomatic}{true}

\newboolean{yMinAutomatic}
\setboolean{yMinAutomatic}{true}
\newboolean{yMaxAutomatic}
\setboolean{yMaxAutomatic}{true}

\newboolean{LegendAutomatic}
\setboolean{LegendAutomatic}{true}

\usepackage[firstpage=true]{background}
\SetBgContents{\textcolor{black}{\footnotesize This work has been submitted to the IEEE for possible publication. Copyright may be transferred without notice, after which this version may no longer be accessible.}}
\SetBgPosition{current page.south}
\SetBgVshift{0.50cm}
\SetBgOpacity{1.0}
\SetBgAngle{0.0}
\SetBgScale{1.0}

\IEEEoverridecommandlockouts

\begin{document}

\title{Cooperative Multi-Task Semantic Communication for Joint Classification and Regression Tasks\thanks{This work was supported by the Deutsche Forschungsgemeinschaft (DFG, German Research Foundation) under Germany's Excellence Strategy – EXC-3036 The Martian Mindset, project number: 533607631}\vspace{-0.5em}
}

\author{
\IEEEauthorblockN{Ahmad Halimi Razlighi\,\orcidlink{0009-0006-3826-832X}, Mohammad Siddiqur Rahman\IEEEauthorrefmark{2}\,\orcidlink{0009-0003-4909-2401}, Maximilian H. V. Tillmann\,\orcidlink{0009-0000-6548-4278},\\ Edgar Beck\,\orcidlink{0000-0003-2213-9727}, and Armin Dekorsy\,\orcidlink{0000-0002-5790-1470}}

\IEEEauthorblockA{Department of Communications Engineering, University of Bremen, Germany}

\IEEEauthorblockA{E-mails:\{halimi, tillmann,  beck, dekorsy\}@ant.uni-bremen.de, \IEEEauthorrefmark{2}E-mail: mrahman@uni-bremen.de\vspace{-1em}}


}

\maketitle

\begin{abstract}
Multi-Task semantic communication (SemCom) prioritizes simultaneous execution of multiple tasks over bit-accurate reconstruction in future intelligent networks. In our prior work~\cite{halimi-letter}, we introduced the cooperative multi-task SemCom (CMT-SemCom) framework, in which the semantic encoder is divided into a common unit (CU) and multiple specific units (SUs) to facilitate cooperative multi-task processing. However, CMT-SemCom has been evaluated on homogeneous classification tasks on simplistic datasets, limiting its applicability to real-world perception systems. In this paper, we extend our CMT-SemCom to jointly handle heterogeneous classification and regression tasks on the complex Cityscapes dataset. We adopt the information maximization (InfoMax) principle so that it accommodates mixed discrete and continuous semantic variables. In particular, we benchmark the proposed framework against independent single-task training, a conventional task-agnostic digital transmission, and single-encoder multi-decoder SemCom. Additionally, we investigate the impact of CU capacity on joint task performance, providing design insights. Extensive evaluations demonstrate that CMT-SemCom significantly outperforms the benchmarks. 
\end{abstract}


\section{Introduction} \label{section:Introduction}

The evolution toward intelligent wireless networks is fundamentally reshaping communication paradigms, shifting the design objective from reliable bit transmission to task-oriented information exchange~\cite{taskcom6g}. In this context, semantic communication (SemCom) has emerged as a promising paradigm that prioritizes the transmission of task-relevant semantic information rather than bit-level accuracy~\cite{Gunduz2022}.

Many practical intelligent systems are required to perform multiple inference tasks simultaneously, like autonomous driving, where semantic segmentation, depth estimation, and scene understanding are jointly executed~\cite{chen2018driving}. Such scenarios motivate multi-task SemCom, in which a single semantic representation is shared among several downstream tasks~\cite{MTL-Sem}. Multi-task learning (MTL) exploits common latent representations across related tasks to improve generalization while reducing computational redundancy~\cite{caruana1997multitask}. Motivated by these advantages, recent studies have incorporated MTL into SemCom using deep neural network (DNN) architectures, including a single-encoder on the transmitter (Tx) side and multiple-decoder on the receiver (Rx) side~\cite{MT-SemSensing,MT-SemSynesthesia}. Nevertheless, these approaches primarily treat the SemCom system as a black-box neural network and rely on heuristic network designs without an explicit information-theoretic objective.

To address this limitation, our previous work introduced the cooperative multi-task SemCom (CMT-SemCom) framework, which establishes a mutual information maximization (InfoMax) foundation for cooperative semantic transmission~\cite{halimi-letter}. Specifically, we modeled a semantic source capable of yielding multiple semantic variables from a shared observation and proposed a split semantic encoder consisting of a common unit (CU) and multiple specific units (SUs) on the Tx. By optimizing an end-to-end (E2E) InfoMax objective through variational approximation, the CU learns semantic features shared across tasks, whereas the SUs preserve task-specific information and channel coding for reliable wireless transmission. Experimental evaluations on homogeneous image classification tasks using the MNIST and CIFAR-10 datasets demonstrated that exploiting statistical dependencies through the CU significantly improves task performance compared with independent task processing~\cite{halimi_ojcoms,halimi_icc}.

Despite these encouraging results, the applicability of the existing CMT-SemCom framework remains limited. Previous evaluations considered only homogeneous image classification tasks on relatively simple datasets, whereas practical intelligent systems commonly involve heterogeneous tasks with fundamentally different statistical characteristics. In particular, classification and regression tasks frequently coexist in perception applications. For example, autonomous driving systems simultaneously perform semantic segmentation, which predicts discrete semantic labels, and depth estimation, which infers continuous geometric information from the same visual scene~\cite{chen2018driving}. Since these tasks originate from a shared observation while exhibiting complementary semantic dependencies, they constitute a natural application scenario for the CMT-SemCom.

Joint learning of classification and regression has been extensively investigated in the computer vision and MTL literature from a machine learning (ML) perspective~\cite{mousavian2016jointClssifReg,zhang2018jointclassifReg,hoyer2021JointClssifReg}. More recently, heterogeneous tasks have also been considered in SemCom through learning-based feature sharing mechanisms~\cite{yu2025multitasksemcom}. However, existing approaches optimize task-specific learning losses directly and therefore lack an information-theoretic formulation that explicitly models the underlying semantic variables and their probabilistic relationships. Consequently, it remains unclear how cooperative semantic representations should be learned when discrete and continuous semantic variables coexist within a unified communication framework. Addressing this gap is essential for evaluating the scalability and practical applicability of CMT-SemCom under realistic multi-task workloads and complex real-world datasets.

In this paper, we extend the CMT-SemCom framework to support the joint extraction of task-relevant information for heterogeneous classification and regression tasks from an information-theoretic perspective by modeling discrete and continuous semantic variables. The proposed framework is evaluated on the real-world Cityscapes dataset~\cite{Cordts2016Cityscapes}, representing realistic urban perception scenarios.
The main contributions of this paper are summarized as follows:
\begin{itemize}
\item Generalizing the CMT-SemCom framework to accommodate both discrete and continuous semantic variables, executing simultaneous classification and regression tasks through a unified framework.
\item Benchmarking the proposed framework against single-task training, a conventional task-agnostic digital transmission pipeline, and single-encoder multi-decoder (SEMD-SemCom).
\item Investigating the impact of CU capacity on multi-task performance, revealing a trade-off between shared representation richness and task-specific interference, and providing design guidelines for optimizing CU dimensionality.
\end{itemize}

\section{Problem Formulation} \label{section.SystemModel}
In this section, we establish a probabilistic model to characterize the semantic source and communication aspects. Based on that, we define the overall CMT-SemCom optimization problem for joint classification and regression tasks.

\subsection{System Model}\label{subsec.system_prob_model}

We consider a multi-task communication system, consisting of a single Tx, multiple Rxs, and a wireless channel in between. The Tx makes an observation $\bm S \in \mathbb R^{H\times W \times C}$, where $H$, $W$, and $C$ represent the height, width, and color channels of an input image, respectively. Then it extracts multiple task-specific information via the split structure, each corresponding to one of $N$ downstream tasks. At Rx $i$, this task-specific information is used to decode a reconstruction of \emph{semantic variable} $\bm Z_i$ to support execution of its task. We denote the tuple $(\bm Z, \bm S)$ as \emph{semantic source}~\cite{halimi-letter}, fully described by its probability density function (pdf) $p(\bm{z}, \bm{s})$, where $\bm{Z}=[\,\bm Z_1\, \bm Z_2\,\cdots\,\bm Z_N]$. Each semantic variable $\bm Z_i$ has the alphabet $\mathcal Z_i$, and as we consider pixel-level semantic perception tasks, semantic variables have the dimensionality of $\bm Z_i \in \mathcal Z_i^{H\times W}$. The alphabet $\mathcal Z_i$ is either continuous with $\mathcal Z_i= \mathbb R$, or discrete with $\mathcal Z_i= \{ \text{`class}_1\text{'},\dots,\text{`class}_F\text{'}\}$.

\begin{figure}[t!]
    \centering
    \hspace*{-1.25cm}
    \resizebox{0.56\textwidth}{!}{\definecolor{CU2_color}{RGB}{204,102,119} 
\definecolor{CU1_color}{RGB}{136,204,238}

\definecolor{CU2_color}{RGB}{255,255,255} 
\definecolor{CU1_color}{RGB}{255,255,255}

\usetikzlibrary{patterns, backgrounds}
\begin{tikzpicture}


\node[
    rectangle,  rounded corners=5pt,
    draw=black,
    minimum width=3.5cm, minimum height=1.45cm,
    align=center, font=\normalsize
  ] (S) at (0,-2cm) {
    \textbf{Semantic Source}\\
    $(\bm{z},\bm s)$\\[1pt]
    $\bm{z}=[\bm z_1 \ \bm z_2 \ \cdots \ \bm z_N]$\\[1pt]
    $\bm s\in\mathbb{R}^{H\times W\times3}$
  };

\node[
    draw,
    rectangle,
    right=0.5 cm of S,
    minimum width=1.5cm,
    minimum height=0.6cm,
    inner sep=0pt
] (C) {
    {CU Enc.} 
};

\node[above=0.1cm of C,xshift=0.7cm](cLabel){$\bm c$};

        \node[draw, rectangle, right=0.3 cm of C, yshift=1.1 cm, inner sep=4pt] (SU1) {$\text{SU Enc.}$};
        \node[draw, rectangle, right=0.3 cm of C, yshift=0.0 cm, inner sep=4pt] (SU2) {$\text{SU Enc.}$};
        \node at ([xshift=1.05cm, yshift=-0.45cm]C.east) {$ \large \vdots$};
        \node[draw, rectangle, right=0.3 cm of C, yshift=-1.1 cm, inner sep=4pt] (SU3) {$\text{SU Enc.}$};
        
        \begin{scope}
            \begin{pgfonlayer}{background}
            \path[pattern=north west lines, pattern color=CU1_color] 
            ([xshift=-0.75cm, yshift=0.27cm]SU1.center) 
            rectangle 
            ([xshift=0.75cm, yshift=-0.26cm]SU1.center); 

            \path[pattern=north west lines, pattern color=CU1_color] 
            ([xshift=-0.75cm, yshift=0.27cm]SU2.center) 
            rectangle 
            ([xshift=0.75cm, yshift=-0.26cm]SU2.center);
            \end{pgfonlayer}
        \end{scope}

        \begin{scope}
            \begin{pgfonlayer}{background}
            \path[pattern=north east lines, pattern color=CU2_color] 
            ([xshift=-0.75cm, yshift=0.27cm]SU3.center) 
            rectangle 
            ([xshift=0.75cm, yshift=-0.26cm]SU3.center); 
            \end{pgfonlayer}
        \end{scope}

        \node[right=0.4 cm of SU1](X1){$\bm{x}_1$};
        \node[right=0.4 cm of SU2](X2){$\bm{x}_2$};
        \node[right=0.4 cm of SU3](Xn){$\bm{x}_N$};

        \renewcommand{\arraystretch}{0.85}
        \node[right=of SU3, yshift=1.05 cm, xshift=-0.02 cm] (ch) {$\begin{tabular}{@{}c@{}}
              \footnotesize C\\
              \footnotesize h\\
              \footnotesize a\\
              \footnotesize n\\
              \footnotesize n\\
              \footnotesize e\\
              \footnotesize l
        \end{tabular}$};
        \node[draw, fit=(ch), minimum height=2.0 cm, inner sep=-1pt]{};
        
        \node[draw, rectangle, right=2.5 of SU1] (Rx1) {$\text{Dec.}$};
        \node[draw, rectangle, right=2.5 of SU2] (Rx2) {$\text{Dec.}$};
        \node[draw, rectangle, right=2.5 of SU3] (Rx3) {$\text{Dec.}$};
        \node at ([xshift=4.75cm, yshift=-0.45cm]C.east) {$ \large \vdots$};
        
        \begin{scope}
            \begin{pgfonlayer}{background}
            \path[pattern=north west lines, pattern color=CU1_color] 
            ([xshift=-0.44cm, yshift=0.23cm]Rx1.center) 
            rectangle 
            ([xshift=0.44cm, yshift=-0.23cm]Rx1.center); 

            \path[pattern=north west lines, pattern color=CU1_color] 
            ([xshift=-0.44cm, yshift=0.23cm]Rx2.center) 
            rectangle 
            ([xshift=0.44cm, yshift=-0.23cm]Rx2.center); 

            \path[pattern=north east lines, pattern color=CU2_color] 
            ([xshift=-0.44cm, yshift=0.23cm]Rx3.center) 
            rectangle 
            ([xshift=0.44cm, yshift=-0.23cm]Rx3.center);
            \end{pgfonlayer}
        \end{scope}
        
        \node[left=0.4 cm of Rx1](X1_hat){$\bm{y}_1$};
        \node[left=0.4 cm of Rx2](X2_hat){$\bm{y}_2$};
        \node[left=0.4 cm of Rx3](Xn_hat){$\bm{y}_N$};
        \node[right=0.4 cm of Rx1](Z1_hat){$\hat{\bm z}_1$};
        \node[right=0.4 cm of Rx2](Z2_hat){$\hat{\bm z}_2$};
        \node[right=0.4 cm of Rx3](Zn_hat){$\hat{\bm z}_N$};


        \draw[->, line width=1pt] (S) -- node[above]{\small$\bm{s}$} (C.west);
        \draw[->, line width=1pt] ([]C.east) -- (SU1.west);
        \draw[->, line width=1pt] ([]C.east) -- (SU2.west);
        \draw[->, line width=1pt] ([]C.east) -- (SU3.west);
        \draw[->, line width=1pt] (SU1) -- (X1);
        \draw[->, line width=1pt] (SU2) -- (X2);
        \draw[->, line width=1pt] (SU3) -- (Xn);
        \draw[->, line width=1pt] (X1_hat) -- (Rx1);
        \draw[->, line width=1pt] (X2_hat) -- (Rx2);
        \draw[->, line width=1pt] (Xn_hat) -- (Rx3);
        \draw[->, line width=1pt] (Rx1) -- (Z1_hat);
        \draw[->, line width=1pt] (Rx2) -- (Z2_hat);
        \draw[->, line width=1pt] (Rx3) -- (Zn_hat);

    \end{tikzpicture}}
    \caption{CMT-SemCom system model.}
    \label{fig:cmt_model}
\end{figure}
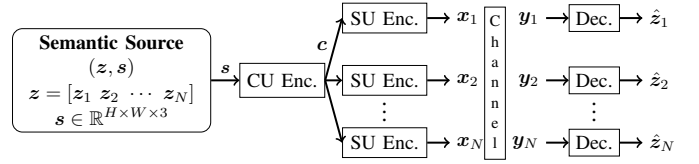


We consider the split encoder structure, consisting of a CU and $N$ SUs. Let $(\bm z, \bm s)$ be a semantic source sample, where only $\bm s$ is available to the Tx. Then, the CU preprocesses this observation $\bm s$ for a subset of semantically related tasks, and the $i$-th SU takes the CU's output $\bm c$ and computes a length-$d$ codeword $\bm x_i$. This codeword is transmitted over an AWGN channel to Rx $i$. The received signal $\bm y_i = \bm x_i + \bm n_i$ with $\bm n_i \sim N(\bm 0_d, \gamma_n^2 \bm I_d)$ is used by decoder $i$ to infer $\hat{\bm z}_i$, with the goal that $\hat{\bm z}_i = \bm z_i$. Thus, we have the joint pdf for the $i$-th semantic link as: 
\begin{equation}
    \begin{aligned}
     p(\bm z_i,\bm{s},\bm{c}_k,\bm{x}_i,\bm{y}_i,\hat{\bm z}_i) =\\[0.0em] & \hspace{-9.8em} p(\bm{s}, \bm z_i)p^{\text{\tiny CU}}(\bm{c}|\bm{s})p^{\text{\tiny SU}}(\bm{x}_i|\bm{c})p(\bm{y}_i|\bm{x}_i)p(\hat{\bm z}_i|\bm{y}_i).
    \end{aligned}
\label{eq:system_probability_markov}
\end{equation}

\subsection{Problem Statement} \label{subsec.Loss_function}
Following the InfoMax principle as in \cite{halimi-letter}, we maximize the mutual information between channel output $\bm Y_i$ and corresponding semantic variable $\bm Z_i$ to design encoders and decoders.

\begin{equation} \label{eq:optimization_problem}
        \left[\,p^\star(\bm{c}|\bm{s}),\, p^\star(\bm{x}_i|\bm{c})\right]\ = \argmax_{p(\bm{c}|\bm{s}), p(\bm{x}_i|\bm{c})} \sum_{i=1}^{N}\, I(\bm{Y}_i; \bm Z_i).
\end{equation}

The optimization variables $p(\bm{c}|\bm{s})$ and $p(\bm{x}_i|\bm{c})$ represent the distributions for the CU and SUs, respectively. For simplicity in notation, we represent $p(\bm{c}|\bm{s})$ and $p(\bm{x}_i|\bm{c})$ by $p^{\text{\tiny CU}}$ and $p^{\text{\tiny SU}}_{i}$, respectively.

\section{CMT-SemCom for Joint Classification and Regression Tasks}

Expanding the mutual information in our optimization problem, as discussed in detail in~\cite{halimi-letter}, the approximated objective function is derived as
\begin{equation} \label{eq:objective_function}
    \begin{aligned}
        &\mathcal{L}(\boldsymbol{\theta}, \boldsymbol{\Phi}, \boldsymbol{\Psi}) = \, \sum_{i=1}^N \, I(\bm{Y}_i; \bm{Z}_i) \\[-0.1cm]
        &\approx 
        \mathbb{E}_{p^{\text{\tiny CU}}_{\boldsymbol{\theta}}}\Bigg[\,\sum_{i=1}^N \Bigg\{ \underbrace{\textcolor{black}{\mathbb{E}_{p(\bm{s},z_i)} \bigg[\,\mathbb{E}_{p^{\text{\tiny SU}}_{\boldsymbol{\phi}_i}} [\,\log q_{\boldsymbol{\psi}_i}(\bm{z}_i|\bm{y}_i)]\ \bigg] } }_{\mathcal L_i(\boldsymbol{\phi_i},\boldsymbol{\psi}_i)} \,\Bigg\} \Bigg].
    \end{aligned}
\end{equation}

In (\ref{eq:objective_function}), we have employed the variational method using weights in neural networks (NNs)~\cite{kingma2013auto}. Thus, our posterior distributions are approximated by NNs, resulting in $p^{\text{\tiny CU}}_{\boldsymbol{\theta}}$ and $p^{\text{\tiny SU}}_{\boldsymbol{\phi}_i}$,  where $\boldsymbol{\theta}$ and $\boldsymbol{\phi}_i$ represents the NN's parameters approximating the CU and $i$-th SU, respectively. We emphasize the joint semantic and channel coding performed by the encoders as we consider $p_{\boldsymbol{\phi}_i}(\bm{y}_i|\bm{s}_i)=\int p^{\text{\tiny SU}}_{\boldsymbol{\phi}_i}(\bm{x}_i|\bm{c})\,p(\bm{y}_i|\bm{x}_i)\,d\bm{x}_i$. For the sake of simplicity we use $p^{\text{\tiny SU}}_{\boldsymbol{\phi}_i} =p_{\boldsymbol{\phi}_i}(\bm{y}_i|\bm{s}_i)$ in (\ref{eq:objective_function}). In addition, although the $i$-th decoder $p(\bm z_i|\bm{y}_i)$ can be fully determined using the known distributions, due to the high-dimensional integrals, we approximate it with $q_{\boldsymbol{\psi}_i}(z_i|\bm{y}_i)$, where $\boldsymbol{\psi}_i$ represents NNs approximating the distribution of the $i$-th decoder.



To obtain the empirical estimate of the above CMT-SemCom objective function, we approximate the expectations using Monte Carlo sampling, considering the training data set of $M$ samples indexed by $m$ \mbox{$\mathcal D = \{ (\bm s^{(m)}, \bm z^{(m)}_1,\dots,\bm z^{(m)}_N) \}_{m = 1}^{M}$}. In the following, we present the empirical loss functions $\mathcal L_i(\boldsymbol{\theta},\boldsymbol{\phi_i},\boldsymbol{\psi}_i)$ for classification tasks and regression tasks separately.

\subsection{Classification Loss Function}
To maximize the objective function \eqref{eq:objective_function}, which is the negative
cross-entropy, we estimate the cross-entropy via data samples $m$ and noise samples $n$, which results in the following maximization objective for the classification tasks:

 \begin{equation} \label{eq:empirical_objective}
    \mathcal{L}_i^\text{Classif}(\boldsymbol{\theta}, \boldsymbol{\phi}_i, \boldsymbol{\psi}_i) \approx   \Biggr. 
          \frac{1}{M}\!\!\sum_{m=1}^{M}\!\bigg[\frac{1}{L}\! \!\sum_{l=1}^{L} [\,\log q_{\boldsymbol{\psi}_i}(\bm{z}_i^{(m)}|\bm{y}_i^{(m,l)})]\bigg] ,
\end{equation}
where we fix the sample size of the encoder output and the channel sampling $\bm{y}_{m,l} = \bm{x}_{m} + \bm{n}_l$, equal for each batch. For implementation, we then minimize the sample-estimated cross-entropy instead of maximizing the negative sample-estimated cross-entropy in \eqref{eq:empirical_objective}.

Finally, for the classification inference step, we take the maximum likelihood of the learned $q^\star_{\boldsymbol{\psi}_i}(\bm{z}_i|\bm{y}_i)$ to get the estimated semantic variable  $\hat{\bm{z}}_i$
\begin{equation}
    \hat{\bm{z}}_i = \argmax_{\hat{\bm{z}}_i}  q^\star_{\boldsymbol{\psi}_i}(\bm{z}_i|\bm{y}_i).
\end{equation}

\subsection{Regression Loss Function} \label{sec:MSE}
Optimizing over $q_{\bm{\psi}_i}(\bm{z}_i|\bm{y}_i)$ for a regression task is generally computationally infeasible, as this amounts to optimizing over an infinite-dimensional space of continuous pdfs. 
A practical way to estimate $q_{\bm{\psi}_i}(\bm{z}_i|\bm{y}_i)$ is to assume a parameterized continuous distribution shape with a tractable number of parameters \cite{wasserman2004all}. Thus, we assume a pixel-wise independent Gaussian pdf with a tractable number of parameters, given by $z_{i_k} \sim \mathcal {N} (\mu_k, \sigma^2)$ for the $k$-th pixel among the $ K=H\times W$ pixels in the input image. We assume equal variances for all pixels' pdfs, and then the \ac{MSE} loss is obtained from \eqref{eq:objective_function} \cite{GoodfellowDeepLearning}.

Assuming a given fix variance $\sigma^2$, the output of $q_{\bm{\psi}_i}(\bm{z}_i|\bm{y}_i)$, which is represented by a NN, is the mean vector $\bm{\mu}_{\bm{\psi}_i}(\bm{y}_i) \in \mathbb R^K$ of the Gaussian distributions, which depends on input $\bm{y}_i$. For simplicity in notation, we omit including the $\bm{\psi}_i$ in $\bm{\mu}(\bm{y}_i)$. 

The pdf $q_{\bm{\psi}_i}(\bm{z}_i|\bm{y}_i)$ is fully described by the mean $\bm{\mu}(\bm{y}_i)$, and the log-likelihood function is given as:
\vadjust{\allowbreak}
\begin{align}    
      & \log q_{\bm{\psi}_i}(\bm{z}_i|\bm{y}_i ) \nonumber \\ 
      & =  \log \left( \prod_{k=1}^K    \frac{1}{\sqrt{2\pi\sigma^2}} \exp  \left( -\frac{  \left[z_{i_k} - \mu_{k}\left(\bm{y}_i\right)\right]^2 \!}{2\sigma^2 } \right)  \right) \nonumber \\[-0.1em] 
    &  =  -   \sum_{k=1}^K   \left(   \frac{\left[z_{i_k} - \mu_{k}\left(\bm{y}_i\right)\right]^2}{2\sigma^2}   \right)      -    \frac{K}{2} \log  \left( 2\pi\sigma^2\right)  .  \label{MSE_derived}
\end{align}

The constant term $\frac{K}{2} \log  \left( 2\pi\sigma^2\right)$ in \eqref{MSE_derived} is ignored in the maximization of the objective function, and we set $K=2\sigma^2$. Integrating \eqref{MSE_derived} into the objective function in \eqref{eq:objective_function}, we obtain the empirical regression loss function as follows:

\begin{equation} \label{eq:regression_empirical_objective}
    \mathcal{L}_i^\text{Regr}(\boldsymbol{\theta}, \boldsymbol{\phi}_i, \boldsymbol{\psi}_i) \approx  \Biggr. 
         -\frac{1}{M}\!\sum_{m=1}^{M}\bigg[\,\frac{1}{L} \!\sum_{l=1}^{L} \,\!\text{MSE} \! \left(\bm{z}_i^{(m)},\bm{\mu}(\bm{y}_i^{(m,l)})\right)\!\bigg] ,
\end{equation}
with \mbox{$\text{MSE}(\bm{z}_i,\bm{\mu}(\bm{y}_i))=\frac{1}{K}\sum_{k=1}^K \! \left(z_{i_k} - \mu_{k}(\bm{y_i})\right)^2$}.

Finally, semantic variable $\hat{\bm{z}}_i$ is estimated using the maximum likelihood criterion by
\begin{equation}
    \hat{\bm{z}}_i = \argmax_{\hat{\bm{z}}_i}  q^\star_{\boldsymbol{\psi}_i}(\bm{z}_i|\bm{y}_i) = \bm{\mu}(\bm{y}_i),
\end{equation}
which is the mean of $q^\star_{\bm{\psi}}(\bm{z}_i|\bm{y}_i )$ as it is Gaussian-shaped, and has its maximum at $\bm{\mu}(\bm{y}_i)$.

\section{Simulation Results} \label{section:Simulation}
In this section, we evaluate the performance of CMT-SemCom by comparing its task execution against baselines, in addition to investigating the impact of the CU capacity.

\subsection{Simulation Setup}

\subsubsection{Dataset and Tasks}

For our evaluations, we use the Cityscapes dataset~\cite{Cordts2016Cityscapes}, which contains $5,000$ finely annotated urban street-scene images collected from $50$ cities. We use only the fine-annotated images, as the dataset provides both pixel-level semantic labels and stereo disparity maps, making it suitable for simultaneous classification and regression tasks. Since disparity is inversely proportional to depth, we convert the disparity maps into normalized floating-point depth labels using the maximum disparity value. Only pixels with positive disparity values are considered valid, while pixels with zero disparity are treated as invalid and excluded from the depth estimation evaluation.

The dataset is shaped as \mbox{$\mathcal D = \{ (\bm s^{(m)}, \bm z^{(m)}_1,\bm z^{(m)}_2,\bm z^{(m)}_3) \}_{m = 1}^{M}$}, with $\bm Z_1\in \mathcal{Z}_1^{H\times W}$, $\bm Z_2\in \mathcal{Z}_2^{H\times W}$, and $\bm Z_3 \in \mathbb R^{H\times W}$. Specifically, the input images we use from the dataset have a resolution of $H\times W \times C = 256\times512\times3$. 


We consider the execution of three tasks $N=3$. For the first two tasks, we consider semantic segmentation of classes: $\mathcal{Z}_1=\{\text{`road', `sidewalk', `building', `car', `otherwise'}\}$ as Task~1 and semantic segmentation of $\mathcal{Z}_2=\{\text{`building', `otherwise'}\}$ as Task~2. Thus, Task~1 is a multinomial classification, and Task~2 is a binary classification. We consider the third task to be a regression task by defining Task~3 to be pixel-level depth estimation, with an illustrative example of the dataset shown in Fig.~\ref{fig:visualization}. Using \eqref{eq:empirical_objective} and \eqref{eq:regression_empirical_objective}, the total loss function of the CMT-SemCom is given as:
\begin{equation*}
    \begin{aligned}
    & \mathcal{L}(\boldsymbol{\theta}, \boldsymbol{\Phi}, \boldsymbol{\Psi}) =\\[-0.1cm] &\mathcal{L}_1^\text{Classif}(\boldsymbol{\theta}, \boldsymbol{\phi}_1, \boldsymbol{\psi}_1) + \mathcal{L}_2^\text{Classif}(\boldsymbol{\theta}, \boldsymbol{\phi}_2, \boldsymbol{\psi}_2) + \mathcal{L}_3^\text{Regr}(\boldsymbol{\theta}, \boldsymbol{\phi}_3, \boldsymbol{\psi}_3).
    \end{aligned}
\end{equation*}



\begin{table}[!t]
\vspace{0.03in}
    \setlength{\tabcolsep}{4pt}
    \centering
    \caption{NN architecture of the CMT-SemCom.}
    \label{table_nn_architectures}
        \centering
        \begin{tabular}{p{0.05\columnwidth}|p{0.8\columnwidth}}
            & \textbf{Layer} \\
            \hline
            \multirow{2}{*}{ \hspace{-1.2mm}\textbf{CU}}  
             & Conv (\#filters $=96$, stride $=2$), BatchNorm, ReLU\\ 
             & $F$ ResNet blocks \\
             \hline 
             \multirow{3}{*}{\textbf{SU}} & $9-F$ ResNet blocks \\
             & Global average pooling 
             \\
             & Dense (output size $=100$), BatchNorm, ReLU
             \\& Dense (output size = number of channel-uses $d=128$), ReLU \\& Power normalization \\
             \hline 
             \multirow{7}{*}{\hspace{-0.2mm}\textbf{Dec}} 
             & Dense, ReLU, Reshape (to $32 \times 64 \times 64$) \\ & ConvT(\#filters $=64$ , stride $=2$), ReLU \\ &ConvT(\#filters $=32$, stride $=2$), ReLU \\ & ConvT(\#filters $=16$, stride $=2$), ReLU \\ &  \textbf{Task 1}: Conv(\#filters $=5$, stride $=\!1$), Softmax
             \\ & \textbf{Task 2}: Conv(\#filters $= 1$ stride $=\!1$), Sigmoid
             \\ &  \textbf{Task 3}: Conv(\#filters $=1$ stride $=\!1$), ReLU \\
             \hline
        \end{tabular}
\end{table}

\subsubsection{CMT-SemCom NN Architecture}

Table~\ref{table_nn_architectures} summarizes the NN architecture of the proposed CMT-SemCom. The CU consists of one convolutional (Conv) layer followed by $F$ ResNet blocks~\cite{ResNet}, while each SU contains the remaining $9-F$ ResNet blocks, keeping the total depth per task fixed at nine blocks. By varying $F$, we investigate how distributing feature extraction between the shared CU and task-specific SUs affects task performance, as discussed in Section~\ref{sub:eval}. The CU output has dimensions $16 \times 32 \times 64$, which is selected after a preliminary investigation to have the best performance. Each SU then applies a global average pooling layer to produce a compact feature vector, which is processed by two fully connected layers and a power normalization layer before transmission with $d$ channel-uses.

The decoder architecture is identical for all tasks except for the final task-specific output layer (Table~\ref{table_nn_architectures}). It first projects the received latent representation through a fully connected layer and reshapes it into a $32 \times 64 \times 64$ feature map. Three transposed convolution (ConvT) layers then progressively upsample the features to the original image resolution of $256 \times 512$, followed by the final task-specific Conv layer.

The network is trained for 120 epochs using the Adam optimizer with an initial learning rate of $10^{-4}$. The learning rate decays by a factor of $10^{-1}$ every 20 epochs starting from epoch 80. 


\definecolor{colorSingle}{RGB}{0,0,0}
\definecolor{colorSU0}{RGB}{0,90,255}
\definecolor{colorSU2}{RGB}{230,120,0}
\definecolor{colorSU4}{RGB}{0,170,0}
\definecolor{colorSU6}{RGB}{230,0,140}
\definecolor{colorSU8}{RGB}{120,85,25}
\definecolor{colorMTNoSU}{RGB}{130,0,140}
\definecolor{colorDigital}{RGB}{30,50,230}

\definecolor{colorSingle}{RGB}{0,0,0}
\definecolor{colorCMTgreen}{RGB}{0,170,0}
\definecolor{colorMTNoSU}{RGB}{230,0,140}

\definecolor{colorSingle}{HTML}{332288} 
\definecolor{colorSU0}{HTML}{88CCEE} 
\definecolor{colorSU2}{HTML}{44AA99} 
\definecolor{colorSU4}{HTML}{117733} 
\definecolor{colorCMTgreen}{HTML}{117733} 

\definecolor{colorSU6}{HTML}{999933} 
\definecolor{colorDigital}{HTML}{DDCC77} 
\definecolor{colorSU6}{HTML}{CC6677} 
\definecolor{colorMTNoSU}{HTML}{882255}  
\definecolor{colorSU8}{HTML}{AA4499} 


\newcommand{\linewidthFigA}{1.0pt}
\newcommand{\linewidthFigB}{2.0pt}
\newcommand{\marksizePlots}{3.5pt}
\newcommand{\marksizePlotsB}{4.0pt}

\newcommand{\markLineWidth}{1.2pt}

\pgfplotsset{
TADplot/.style={
  color=colorDigital,
  line width=\linewidthFigB,
  solid,
  mark=asterisk,
  mark options={solid, fill=white, draw=colorDigital, line width=\markLineWidth},
  mark size=\marksizePlotsB,
  line cap=round,
  line join=round
},
singlePlot/.style={
  color=colorSingle,
  line width=\linewidthFigB,
  solid,
  mark=square,
  mark options={solid, fill=white, draw=colorSingle, line width=\markLineWidth},
  mark size=\marksizePlotsB,
  line cap=round,
  line join=round
},
mtNoSuPlot/.style={
  color=colorMTNoSU,
  line width=\linewidthFigB,
  mark=diamond,
  mark size=\marksizePlotsB,
  mark options={solid, fill=white, draw=colorMTNoSU, line width=\markLineWidth},
  line cap=round,
  line join=round
},
suZeroPlot/.style={
  color=colorSU0,
  line width=\linewidthFigA,
  solid,
  mark=triangle,
  mark repeat=10,
  mark size=\marksizePlots,
  mark options={solid, rotate=90, fill=white, draw=colorSU0, line width=\markLineWidth},
  line cap=round,
  line join=round
},
suTwoPlot/.style={
  color=colorSU2,
  line width=\linewidthFigA,
  solid,
  mark=triangle,
  mark repeat=8,
  mark size=\marksizePlots,
  mark options={solid, rotate=-90, fill=white, draw=colorSU2, line width=\markLineWidth},
  line cap=round,
  line join=round
},
suFourPlot/.style={
  color=colorSU4,
  line width=\linewidthFigA,
  mark=o,
  mark repeat=8,
  mark size=\marksizePlots,
  mark options={solid, fill=white, draw=colorSU4, line width=\markLineWidth},
  line cap=round,
  line join=round
},
cmtGreenPlot/.style={
  color=colorCMTgreen,
  line width=\linewidthFigB,
  mark=o,
  mark size=\marksizePlotsB,
  mark options={solid, fill=white, draw=colorCMTgreen, line width=\markLineWidth},
  line cap=round,
  line join=round
},
suSixPlot/.style={
  color=colorSU6,
  line width=\linewidthFigA,
  mark=x,
  mark repeat=10,
  mark size=\marksizePlots,
  mark options={solid, fill=white, draw=colorSU6, line width=\markLineWidth},
  line cap=round,
  line join=round
},
suEightPlot/.style={
  color=colorSU8,
  line width=\linewidthFigA,
  mark=pentagon,
  mark repeat=10,
  mark size=\marksizePlots,
  mark options={solid, fill=white, draw=colorSU8, line width=\markLineWidth},
  line cap=round,
  line join=round
},
}
\begin{figure*}[t!]
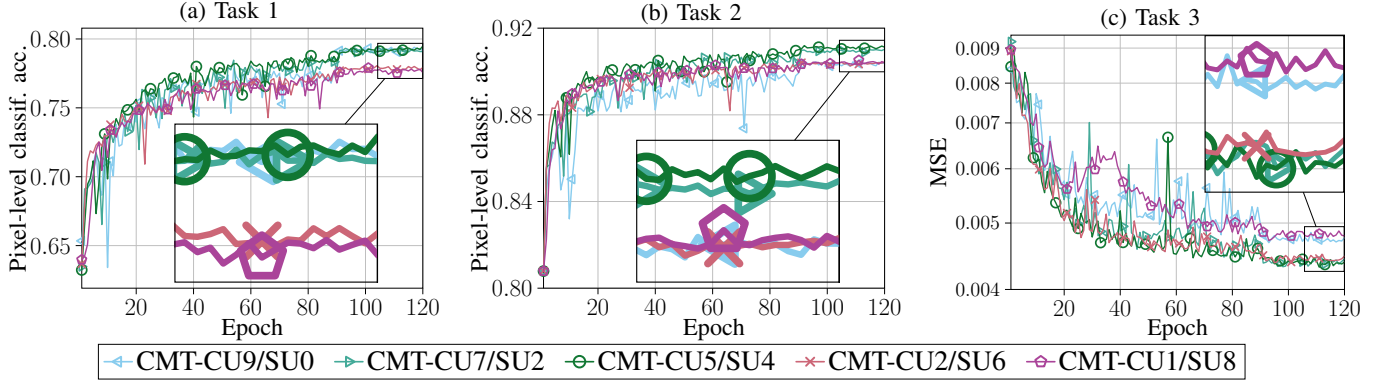

    \centering

    \begin{subfigure}{0.672\columnwidth}
        \centering
        \caption{Task 1}
        \resizebox{\linewidth}{!}{%
            \input{Figures/CombinedSubfigures/Task1_Training.tikz}
        }
        \label{fig:task1_accuracy_a}
    \end{subfigure}
    \hfill
    \begin{subfigure}{0.672\columnwidth}
        \centering
        \caption{Task 2}
        \vspace{-0.16cm}
        \resizebox{\linewidth}{!}{%
            \input{Figures/CombinedSubfigures/Task2_Training.tikz}
        }
        \label{fig:task1_accuracy_b}
    \end{subfigure}
    \begin{subfigure}{0.672\columnwidth}
        \centering
        \caption{Task 3}
        \vspace{0.02cm}
        \resizebox{\linewidth}{!}{%
            \input{Figures/CombinedSubfigures/Task3_Training.tikz}
        }
        \label{fig:task1_accuracy_b}
    \end{subfigure}
    \centering
    \raisebox{-2.6cm}[0pt][0pt]{
    \resizebox{0.85\linewidth}{!}{%
            \begin{tikzpicture}
\begin{axis}[
    hide axis,
    xmin=0, xmax=1,
    ymin=0, ymax=1,
    legend style={
        font=\LARGE,
        draw=black,
        fill=white,
        legend columns=7,
        /tikz/every even column/.append style={column sep=1em}
    },
    legend cell align={left},
]


\addlegendimage{suZeroPlot}
\addlegendentry{CMT-CU9/SU0}

\addlegendimage{suTwoPlot}
\addlegendentry{CMT-CU7/SU2}

\addlegendimage{suFourPlot}
\addlegendentry{CMT-CU5/SU4}

\addlegendimage{suSixPlot}
\addlegendentry{CMT-CU2/SU6}

\addlegendimage{suEightPlot}
\addlegendentry{CMT-CU1/SU8}


\end{axis}
\end{tikzpicture}
        }
        }
    \vspace*{-0.4cm}
    \caption{Architectural comparison  for all three tasks over the training epochs at $0$\,dB SNR.}
\label{fig:arch_comparison}
\end{figure*}

\subsubsection{Benchmarks}

We compare CMT-SemCom's performance against three baselines: single-task SemCom (ST-SemCom), single-encoder multi-decoder SemCom (SEMD-SemCom), and a conventional task-agnostic digital transmission. We note that in all cases, we use the same NN architectural components, e.g., $9$ successive ResNet blocks per task. However, the properties of the NN architectural components are selected such that the total number of trainable parameters remains equal among the cases for a fair comparison.

\textbf{ST-SemCom:} For ST-SemCom, three independent semantic encoder/ decoder are implemented for each task, meaning that there exists no CU shared between them. 

\textbf{SEMD-SemCom:}
We also compare the split structure, CMT-SemCom, against the state-of-the-art MTL in SemCom \cite{MTL-Sem}, where there exists only one shared encoder on the Tx side and multiple decoders on the Rx.

\textbf{Task-agnostic digital transmission (TAD):} 
The benchmark follows a classical separate source and channel coding approach. For source coding, we examined two compression algorithms: JPEG with variable image quality and Huffman coding applied to variable quantization levels of the pixel values. Given our target SNR range and the constrained number of channel uses ($d$), JPEG performed worse than Huffman coding in terms of image reconstruction quality. Therefore, we use Huffman coding as the source coding algorithm. We consider both color and grayscale image reconstruction, with subsampling factors from $1$ to $50$ and quantization levels from $2$ to $8$ bits per pixel, where $8$ bits represents the original pixel precision. Here, subsampling reduces the spatial resolution by retaining fewer pixels, while quantization reduces the number of bits used to represent each pixel.
For channel coding, an LDPC code is used with a rate between $0.2$ and $0.95$. Moreover, modulation schemes of $4$, $16$, and $64$-QAM were examined. All these parameters are optimized by grid search to minimize the MSE of the reconstructed image at $0$\,dB SNR, which resulted in a subsampling factor of $45$, grayscale image reconstruction, quantization to $4$ bits with Huffman coding, a channel coding rate of $0.254$, and the use of $4$-QAM.
Finally, the reconstructed image is fed into a NN to execute all three tasks, using an architecture similar to that of CMT-SemCom. The NN for the task execution is also trained using the noisy reconstructed images for the best possible inference result.

\subsubsection{Evaluation Metrics} 

We evaluate the segmentation tasks by two metrics. First, we consider pixel-level classification accuracy of classes averaged over all image pixels, measuring the fraction of valid pixels whose inferred labels match the corresponding ground-truth labels. Second, we use intersection over union (IoU), which is assessed per class and measures the intersection of the inferred segmentation and the ground truth, divided by the union of true and false positives and false negatives~\cite{IoU}. True positives (TP) are correctly identified object pixels, while false positives (FP) are pixels incorrectly labeled as the object. False negatives (FN) are pixels incorrectly labeled as no object. IoU is defined as:
\begin{equation*}
 \mathrm{IoU} =
\frac{\mathrm{TP}}
{\mathrm{TP}
+\mathrm{FP}
+\mathrm{FN}}.   
\end{equation*}

Particularly, we use IoU to evaluate Task~2 of `building' segmentation, and for Task~1, we use mean IoU (mIoU), which averages the IoU scores for all the defined classes in $\mathcal Z_1$. 







In addition to the MSE, regression performance is evaluated using the $\delta_1$ accuracy \cite{delta1}. $\delta_1$ measures the percentage of predicted depth pixels whose relative error falls within a specific threshold. Specifically, it is the fraction of pixels for which the maximum ratio between the ground-truth depth and the predicted depth is less than $1.25$. 
\begin{equation*}
 \delta_1
=
\frac{1}{K}
\sum_{k=1}^{K}
\mathbb{I} \bigg\{ \max
\bigg(
\frac{z_{k}}
{\hat{z}_{k}},
\frac{\hat{z}_{k}}
{z_{k}} < 1.25 \bigg) \bigg\},   
\end{equation*}
where $\mathbb{I}(\cdot)$ denotes the indicator function and $K$ is the total number of pixels with valid depth ground truth.


\subsection{Evaluations} \label{sub:eval}

\subsubsection{Architectural Investigations}

First, we investigate the impact of the CU/SU processing split on the balance between shared semantic extraction and task-specific refinement. Fig.~\ref{fig:arch_comparison} shows the task performance over training epochs at $0$\,dB SNR and $d=128$ channel-uses for different CMT-SemCom architectures, ranging from CMT-CU9/SU0 to CMT-CU1/SU8. For instance, CMT-CU9/SU0 has all $9$ ResNet blocks in the CU, meaning that most processing is done in the CU with only dense layers in the SU, while CMT-CU1/SU8 shifts most processing to the task-specific SUs with $1$ ResNet block in the CU, and $8$ ResNet blocks in each SU.
As shown in Fig.~\ref{fig:arch_comparison}, a balanced distribution of the processing capacity between the CU and the SU provides the best performance, with CMT-CU5/SU4 achieving the best for all three tasks. Therefore, CMT-CU5/SU4 is adopted as the default CMT-SemCom architecture in the remaining experiments.

\newcommand{\heightTAD}{9.5cm}
\begin{figure}[t!]
    \centering

    \begin{subfigure}{0.492\columnwidth}
        \centering
        \caption{Task 1}
        \vspace{-0.04cm}
        \resizebox{\linewidth}{!}{%
            \begin{tikzpicture}

\begin{axis}[
    ymin=0.580, 
    ymax=0.805,
    xmin=-5,
    xmax=10,
    height=\heightTAD,
    width=8cm,
    ylabel={Pixel-level classif. acc.},
    title style={
        at={(0.5,1.10)},
        anchor=north,
        font=\LARGE
    },
    legend style={
        font=\LARGE,
        draw=black,
        fill=white,
        at={(0.97,0.05)},
        anchor=south east
    },
    ylabel style={font=\LARGE, yshift=0.5em,xshift=-0.4cm},
    xlabel style={font=\LARGE, yshift=+0.2em},
    tick label style={font=\Large},
    yticklabel style={font=\Large,
    /pgf/number format/fixed,
    /pgf/number format/precision=2,
    /pgf/number format/zerofill},
    xlabel ={SNR [dB]},
    xtick={-5,0,5,10},
    grid=major,
    major grid style={solid, gray!30},
  legend style={
    font=\LARGE, draw=black, fill=white,
    at={(0.99,0.5)}, anchor=south east
  },
]

\addplot[cmtGreenPlot] coordinates {
  (-5,0.7567)(0,0.7912)(5,0.7992)(10,0.8011)
};
\addlegendentry{CMT-SemCom}
\addplot[mtNoSuPlot] coordinates {
  (-5,0.7559)(0,0.7913)(5,0.7996) (10,0.8017)
};
\addlegendentry{SEMD-SemCom}
\addplot[singlePlot] coordinates {
  (-5,0.7529)(0,0.7823)(5,0.7903)(10,0.7924)
};
\addlegendentry{ST-SemCom}

\addplot[TADplot] coordinates { 
(-5,0.5836)(0,0.6731)(5,0.6741)(10,0.6741) 
}; \addlegendentry{TAD}


\end{axis}

\end{tikzpicture}
        }
        \label{fig:task1_accuracy_a}
    \end{subfigure}
    \hfill
    \begin{subfigure}{0.492\columnwidth}
        \centering
        \caption{Task 1}
        \vspace{-0.12cm}
        \resizebox{\linewidth}{!}{%
            \begin{tikzpicture}

\begin{axis}[ ymin=0.30, ymax=0.550, xmin=-5, xmax=10,
height=\heightTAD, width=8cm, 
ylabel={mIoU}, title style={ at={(0.5,1.10)}, anchor=north, font=\LARGE },
legend style={ font=\LARGE, draw=black, fill=white, at={(0.97,0.05)}, anchor=south east }, 
ylabel style={font=\LARGE, yshift=0.5em}, xlabel style={font=\LARGE, yshift=+0.2em}, 
tick label style={font=\Large},
yticklabel style={font=\Large, /pgf/number format/fixed, /pgf/number format/precision=2, /pgf/number format/zerofill}, 
xlabel ={SNR [dB]}, 
xtick={-5,0,5,10}, 
grid=major, major grid style={solid, gray!30}, 
legend style={ font=\LARGE, draw=black, fill=white, at={(0.99,0.39)}, anchor=south east },
]

\addplot[cmtGreenPlot] coordinates { (-5,0.4693)(0,0.5199)(5,0.5340)(10,0.5376) }; \addlegendentry{CMT-SemCom}

\addplot[mtNoSuPlot] coordinates { (-5,0.4669)(0,0.5177)(5,0.5320) (10,0.5358) }; \addlegendentry{SEMD-SemCom}

\addplot[singlePlot] coordinates { (-5,0.4630)(0,0.5076)(5,0.5217)(10,0.5259) }; \addlegendentry{ST-SemCom} 

\addplot[TADplot] coordinates { (-5,0.3057)(0,0.3850)(5,0.3862)(10,0.3862) }; \addlegendentry{TAD}


\end{axis}

\end{tikzpicture}
        }
        \label{fig:task1_accuracy_b}
    \end{subfigure}
    
    \vspace*{-0.5cm}
    \begin{subfigure}{0.492\columnwidth}
        \centering
        \caption{Task 2}
        \vspace{-0.12cm}
        \resizebox{\linewidth}{!}{%
            \begin{tikzpicture}

\begin{axis}[
    ymin=0.785, ymax=0.92,
    xmin=-5,
    xmax=10,
    height=\heightTAD,
    width=8cm,
    ylabel={Pixel-level classif. acc.},
    title style={
        at={(0.5,1.10)},
        anchor=north,
        font=\LARGE
    },
    legend style={
        font=\LARGE,
        draw=black,
        fill=white,
        at={(0.97,0.05)},
        anchor=south east
    },
    ylabel style={font=\LARGE, yshift=0.5em,xshift=-0.4cm},
    xlabel style={font=\LARGE, yshift=+0.2em},
    tick label style={font=\Large},
    yticklabel style={font=\Large,
    /pgf/number format/fixed,
    /pgf/number format/precision=2,
    /pgf/number format/zerofill},
    xlabel ={SNR [dB]},
    xtick={-5,0,5,10},
    grid=major,
    major grid style={solid, gray!30},
  legend style={
    font=\LARGE, draw=black, fill=white,
    at={(0.97,0.05)}, anchor=south east
  },
]

\addplot[cmtGreenPlot] coordinates {
  (-5,0.8940)(0,0.9114)(5,0.9139)(10,0.9144)
};
\addplot[mtNoSuPlot] coordinates {
  (-5,0.8922)(0,0.9057)(5,0.9078) (10,0.9082)
};
\addplot[singlePlot] coordinates {
  (-5,0.8924)(0,0.9063)(5,0.9086)(10,0.9092)
};

\addplot[TADplot] coordinates { (-5,0.7872)(0,0.8372)(5,0.8379)(10,0.8379) }; 


\end{axis}

\end{tikzpicture}
        }
        \label{fig:task1_accuracy_a}
    \end{subfigure}
    \hfill
    \begin{subfigure}{0.492\columnwidth}
        \centering
        \caption{Task 2}
        \vspace{0.01cm}
        \resizebox{\linewidth}{!}{%
            \begin{tikzpicture}

\begin{axis}[
    ymin=0.29, ymax=0.615,
    xmin=-5,
    xmax=10,
    height=\heightTAD,
    width=8cm,
    ylabel={IoU},
    title style={
        at={(0.5,1.10)},
        anchor=north,
        font=\LARGE
    },
    legend style={
        font=\LARGE,
        draw=black,
        fill=white,
        at={(0.97,0.05)},
        anchor=south east
    },
    ylabel style={font=\LARGE, yshift=0.5em},
    xlabel style={font=\LARGE, yshift=+0.2em},
    tick label style={font=\Large},
    yticklabel style={font=\Large,
    /pgf/number format/fixed,
    /pgf/number format/precision=2,
    /pgf/number format/zerofill},
    xlabel ={SNR [dB]},
    xtick={-5,0,5,10},
    grid=major,
    major grid style={solid, gray!30},
  legend style={
    font=\LARGE, draw=black, fill=white,
    at={(0.97,0.05)}, anchor=south east
  },
]

\addplot[cmtGreenPlot] coordinates {
  (-5,0.4912)(0,0.5834)(5,0.6020)(10,0.6068)
};
\addplot[mtNoSuPlot] coordinates {
  (-5,0.4903)(0,0.5644)(5,0.5809)(10,0.5851)
  };
\addplot[singlePlot] coordinates {
  (-5,0.4892)(0,0.5635)(5,0.5799)(10,0.5844)
};

\addplot[TADplot] coordinates { (-5,0.2933)(0,0.3760)(5,0.3774)(10,0.3774) }; 


\end{axis}

\end{tikzpicture}
        }
        \label{fig:task1_accuracy_b}
    \end{subfigure}

    \vspace*{-0.5cm}
    \begin{subfigure}{0.492\columnwidth}
        \centering
        \caption{Task 3}
        \vspace{-0.12cm}
        \resizebox{\linewidth}{!}{%
            \begin{tikzpicture}

\begin{axis}[
    ymin=0.00390, ymax=0.010,
    xmin=-5,
    xmax=10,
    height=\heightTAD,
    width=7.8cm,
    ylabel={MSE},
    title style={
        at={(0.5,1.10)},
        anchor=north,
        font=\LARGE
    },
    legend style={
        font=\LARGE,
        draw=black,
        fill=white,
        at={(0.97,0.95)},
        anchor=south east
    },
        log basis y={10},
    ymode=log,
    ymajorgrids,
    yminorgrids,
    minor y tick num=1,
ytick={0.003,0.004,0.005,0.006,0.007,0.008,0.009,0.01,0.012,0.014,0.02,0.03,0.04,0.05,0.06,0.07,0.08,0.09,0.1,0.2,0.3,0.4,0.5,0.6,0.7,0.9,1},
 yticklabels={
  \(\displaystyle {0.003}\),\(\displaystyle {0.004}\),\(\displaystyle {0.005}\),\(\displaystyle {0.006}\),\(\displaystyle {0.007}\),\(\displaystyle {0.008}\),\(\displaystyle {0.009}\),\(\displaystyle {0.010}\),\(\displaystyle {0.012}\),\(\displaystyle {0.014}\),\(\displaystyle {0.02}\),,,\(\displaystyle {0.05}\),,,,,
  \(\displaystyle {0.1}\),\(\displaystyle {0.2}\),\(\displaystyle {0.3}\),\(\displaystyle {0.4}\),\(\displaystyle {0.5}\),,,,,
  \(\displaystyle {10^{0}}\),,,,,,,,,
  \(\displaystyle {10^{1}}\)
},
    ylabel style={font=\LARGE, yshift=+1.6em},
    xlabel style={font=\LARGE, yshift=+0.2em},
    tick label style={font=\Large},
    scaled y ticks=false,
    yticklabel style={
    font=\Large,
    /pgf/number format/fixed,
    /pgf/number format/precision=4,
    /pgf/number format/zerofill
    },
    xlabel ={SNR [dB]},
    xtick={-5,0,5,10},
    grid=major,
    major grid style={solid, gray!30},
  legend style={
    font=\LARGE, draw=black, fill=white,
    at={(0.97,0.05)}, anchor=south east
  },
]

\addplot[cmtGreenPlot] coordinates {
  (-5,0.00554)(0,0.00438)(5,0.00422)(10,0.00418)
};
\addplot[mtNoSuPlot] coordinates {
  (-5,0.00583)(0,0.00492)(5,0.00478)(10,0.00475)
};
\addplot[singlePlot] coordinates {
  (-5,0.00563)(0,0.00455)(5,0.00437)(10,0.00433)
};

\addplot[TADplot] coordinates { (-5,0.00992)(0,0.00658)(5,0.00653)(10,0.00653) }; 


\end{axis}

\end{tikzpicture}
        }
        \label{fig:task1_accuracy_a}
    \end{subfigure}
    \hfill
    \begin{subfigure}{0.492\columnwidth}
        \centering
        \caption{Task 3}
        \vspace{-0.12cm}
        \resizebox{\linewidth}{!}{%
            \begin{tikzpicture}

\begin{axis}[
    ymin=0.49, ymax=0.640,
    xmin=-5,
    xmax=10,
    height=\heightTAD,
    width=8cm,
    ylabel={$\delta_1$ accuracy},
    title style={
        at={(0.5,1.10)},
        anchor=north,
        font=\LARGE
    },
    legend style={
        font=\LARGE,
        draw=black,
        fill=white,
        at={(0.97,0.05)},
        anchor=south east
    },
    ylabel style={font=\LARGE, yshift=0.5em},
    xlabel style={font=\LARGE, yshift=+0.2em},
    tick label style={font=\Large},
    yticklabel style={font=\Large,
    /pgf/number format/fixed,
    /pgf/number format/precision=2,
    /pgf/number format/zerofill},
    xlabel ={SNR [dB]},
    xtick={-5,0,5,10},
    grid=major,
    major grid style={solid, gray!30},
  legend style={
    font=\LARGE, draw=black, fill=white,
    at={(0.97,0.05)}, anchor=south east
  },
]

\addplot[cmtGreenPlot] coordinates {
  (-5,0.5882)(0,0.6201)(5,0.6260)(10,0.6276)
};
\addplot[mtNoSuPlot] coordinates {
  (-5,0.5749)(0,0.6042)(5,0.6100)(10,0.6115)
};
\addplot[singlePlot] coordinates {
  (-5,0.5807)(0,0.6095)(5,0.6152)(10,0.6169)
};

\addplot[TADplot] coordinates {
  (-5,0.4972)(0,0.5655)(5,0.5664)(10,0.5664)
};


\end{axis}

\end{tikzpicture}
        }
        \label{fig:task1_accuracy_b}
    \end{subfigure}
    \vspace*{-0.9cm}
    \caption{Task execution comparison of CMT-SemCom vs. benchmarks over SNR.}
    \label{fig:benchmark_comparison}
\end{figure}
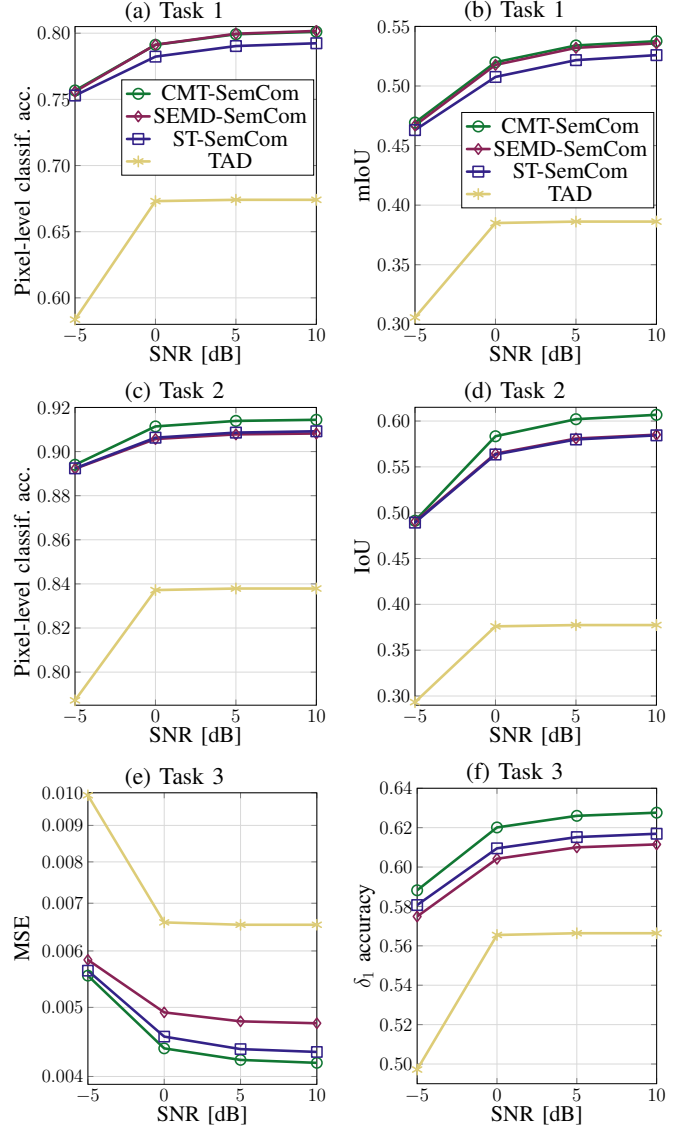


    



\subsubsection{Benchmark Comparisons}
We now compare CMT-SemCom against ST-SemCom, SEMD-SemCom, and TAD over an SNR range as shown in Fig.~\ref{fig:benchmark_comparison}. The training SNR is $0\,\text{dB}$. The comparison uses pixel-level classification accuracy and IoU for Tasks~1 and Task~2, and MSE and $\delta_1$ accuracy for Task~3. Increasing the SNR from $-5$ to $10$\,dB improves performance across all metrics for the considered methods due to the more favorable channel conditions.


The proposed CMT-SemCom framework outperforms ST-SemCom on all three tasks and all SNR values, indicating that joint processing in the CU helps in extracting the relevant semantic feature for all three tasks.

Although SEMD-SemCom achieves the same pixel-level classification accuracy as CMT-SemCom on Task 1, CMT-SemCom attains a slightly higher IoU. However, for Tasks~2 and~3, CMT-SemCom consistently outperforms SEMD-SemCom. These results indicate that the SUs are crucial for task-specific refinements on the Tx side.

Finally, we compare CMT-SemCom with TAD transmission. For a fair comparison, we set the number of channel-uses to $384=3\times128$ in TAD. This keeps the comparison symbol-wise fair between these methods as we transmit $3\times128$ real-symbols for CMT-SemCom and $3\times128$ complex-symbols for TAD. The task execution accuracy after TAD transmission is worse compared to CMT-SemCom, as the image reconstruction quality is low for the limited number of channel-use and $0\,\text{dB}$ SNR. The results highlight the advantage of CMT-SemCom over conventional task-agnostic digital transmission in task-oriented communication, especially in bandwidth-constrained and low-SNR conditions.

\begin{figure}[!t]
  \centering
  \resizebox{0.99\columnwidth}{!}{%
  \input{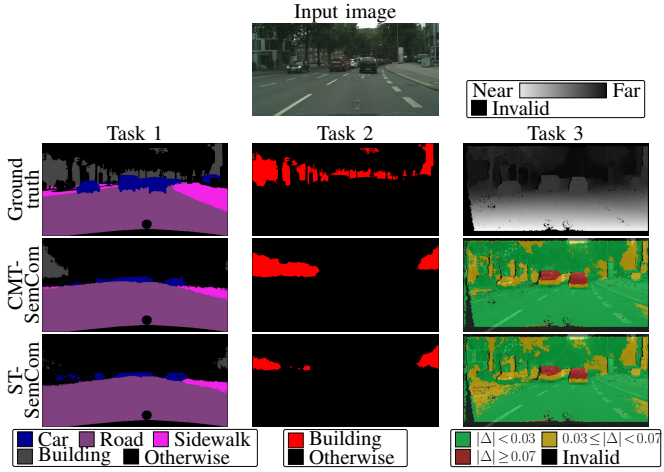}
}
  \caption{Visualization of CMT-SemCom performance on the Cityscapes dataset trained and evaluated at $0$\,dB SNR.}
  \label{fig:visualization}
\end{figure}



Fig.~\ref{fig:visualization} visualizes the performance of CMT-SemCom for all three tasks and compares it with ground truth and ST-SemCom for one illustrative example of the dataset. For both segmentation tasks, as shown, target classes are detected more accurately than by ST-SemCom, and the detected class masks are less fragmented and show better overlap with the ground truth. Regarding Task~3 for better visualization, the mean absolute error $\Delta$ between $\bm z_3$ and the inferred $\hat{\bm{z}}_3$ are shown, where $\bm z_3$ and $\hat{\bm{z}}_3$ are normalized to be between $0$ and $1$. 
Low-error regions with $0\leq \Delta < 0.03$ are indicated in green, moderate-error regions with $0.03 \leq \Delta < 0.07$ are indicated in yellow, high-error regions with $0.07 \leq \Delta \leq 1 $ are indicated in red, and pixels with invalid ground truth are indicated in black. It can be seen that CMT-SemCom has larger low-error regions, less moderate-error regions, and nearly the same high-error regions compared to ST-SemCom.

\section{Conclusion} \label{section:Conclusion}
We extended CMT-SemCom to heterogeneous multi-task perception by jointly supporting classification and regression tasks on the Cityscapes dataset using an InfoMax formulation that accommodates mixed discrete and continuous semantic variables. Simulation results showed that the proposed framework outperforms single-task SemCom, single-encoder multi-decoder SemCom, and task-agnostic digital transmission, while requiring the same communication resources. In addition, we demonstrated that the distribution of feature extraction between the shared CU and task-specific SUs plays a crucial role in balancing information sharing and task specialization, providing practical design insights.

\balance
\bibliographystyle{IEEEtran}
\bibliography{IEEEabrv,References.bib}

\end{document}